\documentclass[12pt,letterpaper,copyedit]{iopart}
\usepackage{epsfig}
\def\be{\begin{equation}}
\def\ee{\end{equation}}
\def\bea{\begin{eqnarray}}
\def\eea{\end{eqnarray}}

\begin{document}
\title{Eigenmode in a misaligned triangular optical cavity }
\author{F.Kawazoe$^1$, R. Schilling$^2$ and H. L\"uck$^1$}

\address{$^1$\,Max-Planck-Institut f\"ur Gravitationsphysik
(Albert-Einstein-Institut) und Leibniz Universit\"at Hannover, Callinstr.~38,
   D--30167 Hannover, Germany }
\address{$^2$\,Max-Planck-Institut f\"ur Gravitationsphysik
(Albert-Einstein-Institut), Fr\"ottmaninger Weg~18,
   D-85748 Garching, Germany}
\ead{fumiko.kawazoe@aei.mpg.de}

\begin{abstract}
We derive relationships between various types of small misalignments 
on a triangular Fabry-Perot cavity and associated geometrical eigenmode
changes. 
We focus on the changes of beam spot positions on cavity mirrors, 
the beam waist position, and its angle. 
A comparison of analytical and 
numerical results shows excellent agreement. 
The results are applicable to any triangular cavity close to an isosceles 
triangle, with the lengths of two sides much bigger than the other, 
consisting of a curved mirror and two flat mirrors yielding a waist 
equally separated from the two flat mirrors. 
This cavity shape is most commonly used in laser interferometry. 
The analysis presented here can easily be extended to more 
generic cavity shapes. 
The geometrical analysis not only serves as a method of checking a simulation
result, but also gives an intuitive and handy tool to visualize the eigenmode
of a misaligned triangular cavity. \end{abstract}


\section{Introduction}
Fabry-Perot cavities are widely used in the field of laser interferometry, 
and longitudinal length shifts of a cavity mirror and the resulting change 
in the phase of the resonating field is well known. 
However in the case where suspended mirrors are used, such as in 
gravitational wave detectors, angular shifts play a crucial role 
in the detector performance; they ensure clean length control signals. 
Angular shifts on the cavity mirrors and resulting eigenmode changes 
in the circulating Gaussian beam of a plane cavity were geometrically 
analyzed in~\cite{Gerhard}, and the results are used, together with 
results from simulation work,  to obtain error signals to control 
the alignment of various cavity mirrors. 
Recently we designed a triangular optical cavity for the purpose of 
frequency stabilization for the AEI~10\,m Prototype\,\cite{Fumiko}, 
and in the process of designing an alignment control system, a 
geometrical analysis for this cavity was performed. 
The cavity is close to an isosceles 
triangle, with the lengths of two sides much bigger than the other, 
consisting of a curved mirror and two flat mirrors yielding a waist 
equally separated from the two flat mirrors. 
However this cavity shape is most commonly used in laser 
interferometry, the results presented here can easily be extended 
to more generic cavity shapes.  
In this paper we first derive the relations of small mirror misalignments 
and the resulting changes in the eigenmode. 
By small misalignments we mean the regime where a lateral shift, 
and the angular deviation of the waist is smaller than the waist radius, 
and the divergence angle of the beam, respectively. 
The results are shown in terms of beam spot position changes on all the 
cavity mirrors, the waist position changes, and the waist angular shifts. 
They carry sufficient information for designing an angular control system. 
We then compare the results with that of two simulation tools, and show that 
they are in excellent agreement with each other.

\section{Types of misalignments}
\begin{figure}[htbp]
\centering
\includegraphics[width=120mm]{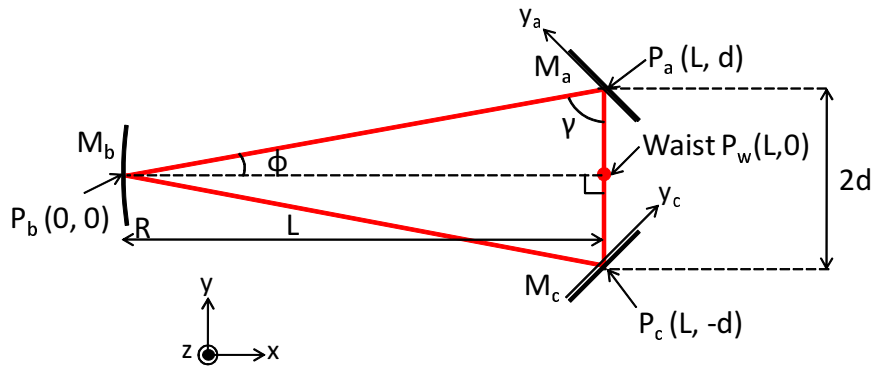}
\caption{Schematic of an aligned triangular cavity within the $x-y$ plane.
Also defined are the two coordinate axes $y_a$ and $y_c$ that are fixed on the
flat mirrors $\rm M_a$ and $\rm M_c$, respectively. Mirror $\rm M_b$ has a
radius of curvature of~$R$.}
\label{000}
\end{figure}

Figure~\ref{000} shows the schematic of a triangular cavity when aligned. 
Two flat mirrors are relatively close together and are
labeled as $\rm M_a$ and $\rm M_c$, while the curved mirror is far away,
has a radius of curvature~R and is labeled as~$\rm M_b$. 
The position where the beam hits the mirror $\rm M_i$ is given by $P_i$, 
as well as the waist position by $P_w$, followed by the associated 
coordinates within the $x-y$ plane.
Here, we also introduce a coordinate system attached to each of the flat
mirrors ($y_a$ and $y_c$) for convenience. 
The two equal angles of the beam at $\rm M_a$ and $\rm M_c$ and half the 
small angle at $\rm M_b$ are given by $\gamma$
and $\phi$, respectively. 
Due to the shape of the triangle the following approximations hold and are
used throughout this paper unless otherwise noted:
\begin{equation}\label{gamap}
\gamma \approx\pi/2
\end{equation}
\vspace{-4mm}
\begin{equation}\label{phiap}
\phi\ll 1
\end{equation}
Angular degrees of freedom in horizontal and vertical directions for the
three mirrors produce six modes of misalignments. 

Misalignment angles of mirror $\rm M_i$ are given by $\alpha_i$ and 
$\beta_i$ for horizontal (angles around the $z$-axis, sometimes also
called yaw or rotation), and vertical 
(inclination angle with respect to the $x-y$ plane, sometimes also
called pitch or tilt) directions, respectively. 
A positive angle is formed by counter-clockwise rotation around the $z$-axis
for horizontal misalignments, and around the $y$-axis, $y_a$-axis, and
$y_c$-axis for vertical misalignment of $\rm M_b$, $\rm M_a$ and $\rm M_c$,
respectively.
We take linear combinations of these two flat mirror misalignments to form
common and differential modes: $\alpha_\pm=\left(\alpha_a\pm\alpha_c\right)/2$
and $\beta_\pm=\left(\beta_a\pm\beta_c\right)/2$. 
The changes in the waist position and the beam spot position on mirror $\rm
M_i$ are denoted by $\Delta k_w$ and $\Delta k_i$, with $k$ being the
corresponding $x$ or $y$ coordinates. 
Ane angular change of the beam between the two flat mirrors is denoted by 
$\theta$ (see Fig.~\ref{003}. 
Since we concern small misalignments, these changes are also small. 
Hence we use the following approximation throughout this paper: $\theta \ll 1$
and $O\left(\Delta k^n\right)=0$ for $n\geq 2$.
All types of misalignments are summarized and the associated section numbers are
listed in Table~\ref{types}.
\small
\begin{table}
\caption{\label{types}Summary of types of misalignments and associated section
numbers.} 
\begin{indented}
\item[]\begin{tabular}{@{}llll}
\br
Type&Description&Section\\
\mr
$\alpha_-$&Differential of the flat mirrors in horizontal &$3.1.$\\
$\alpha_b$&Curved mirror in horizontal &$3.2.$\\
$\alpha_+$&Common of the flat mirrors in horizontal &$3.3.$\\
$\beta_b$&Curved mirror in vertical &$4.1.$\\
$\beta_+$&Common of the flat mirrors in vertical &$4.2.$\\
$\beta_+$&Differential of the flat mirrors in vertical &$4.3.$\\
\br
\end{tabular}
\end{indented}
\end{table}
\section{Horizontal misalignments}
\subsection{Misalignment in $\alpha_-$}
\begin{figure}[htb]
\centerline{\includegraphics[width=120mm]{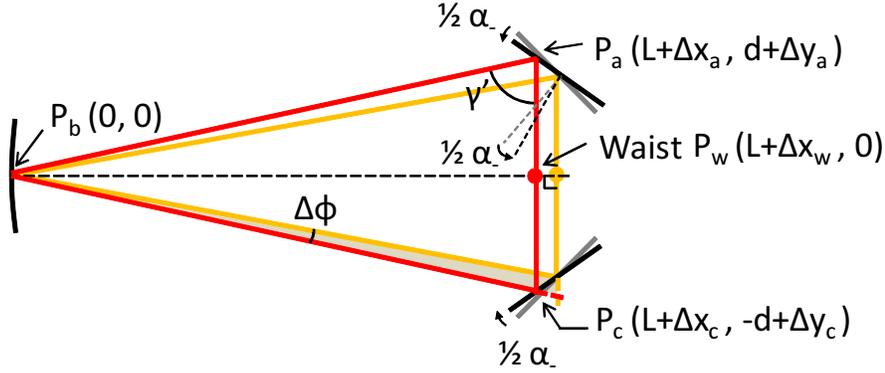}}
\caption{Cavity eigenmodes of the aligned (lighter colored triangle) and the
misaligned by $\alpha_-$ (darker colored triangle) cases. The contrary tilts 
around the $z$-axis cause a symmetric change in the eigenmode.\label{001}}
\end{figure}

A misalignment in $\alpha_-$, i.e.\ contrary tilts around the $z$-axis, 
keep the cavity symmetric to the $x$-axis and, hence, causes a symmetric 
change in the eigenmode. 
In Fig.~\ref{001}, the original and the new eigenmodes are shown by the
lighter (yellow) and darker colors (this color rule is applied 
throughout this paper), and the $x$ and $y$ coordinates of the spot positions
on the mirrors are shown.
Because of the symmetry it is obvious that $\Delta x_a$
equals $\Delta x_c$ and $\Delta x_w$, and due to the approximation given by
Equ.~\ref{gamap}, $\Delta y_a$ also equals $\Delta x_a$.
The angle of incidence on the flat mirrors changes by $\frac{1}{2} \alpha_-$,
as indicated by the dashed normal on one mirror surface. 
The large angle $\gamma^{'}$ changes by $-\alpha_-$, 
yielding a change by $\alpha_-$ in half the small angle $\Delta\phi$. 
From looking at the shaded area in Fig.~\ref{001} we get:
\begin{equation}
\Delta y_c \approx\,\sqrt{L^2+d^2}\,\sin{\Delta \phi}\approx\,\sqrt{L^2+d^2}\cdot\Delta \phi=\,-\sqrt{L^2+d^2}\cdot\alpha_-
\end{equation}
Therefore we know the following relations between the spot position changes
and the misalignment angle: 
\begin{equation}
\Delta x_a =\,\Delta x_c =\,-\Delta y_a =\,\Delta y_c =\,\Delta x_w=\,
-\sqrt{L^2+d^2}\cdot\alpha_- 
\end{equation}
and hence the angle deviation $\theta$ at the waist is zero.

To summarize, a misalignment in $+(-)\alpha_-$ causes a shrink (stretch)
of the eigenmode along the $x$-axis, yielding the eigenmode to keep its 
isosceles shape, but change its shape in a way that it becomes more
``fat'' (``thin''). 
As a result, the waist position shifts in $x$-direction by an amount that is
approximately proportional to the distance between the 
curved mirror and the two flat mirrors.

\subsection{Misalignment in $\alpha_b$}
\begin{figure}[htb]
\centerline{\includegraphics[width=120mm]{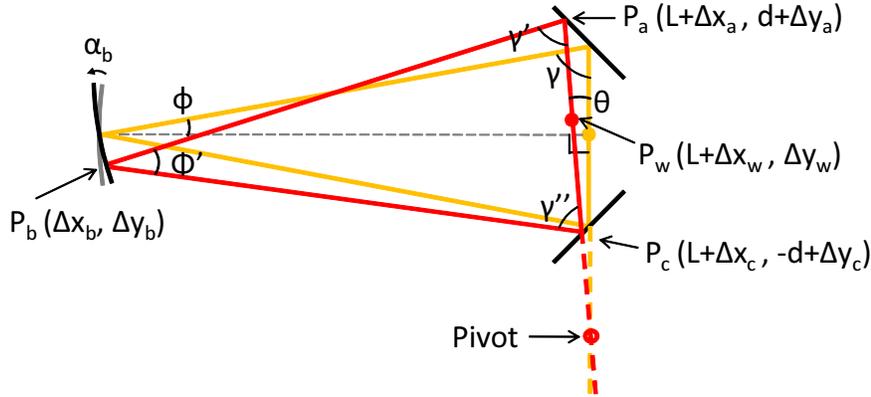}}
\caption{Cavity eigenmodes of the aligned and the misaligned cases by
$\alpha_b$. We start with a general, and hence hypothetical case where the
pivot and the bisecting point of the non-congruent do not match, and later
show they coincide. The changes of the two larger angles ($\gamma$)
are of the same size but with opposite sign, hence the small
angle $\phi$ stays unchanged.\label{003}}
\end{figure}

Figure~\ref{003} shows a hypothetic misaligned cavity caused by~$\alpha_b$,
i.e.\ a rotation of $\rm M_b$ around the vertical axis.
In this case, there is no obvious symmetry axis. 
One can expect changes in the positions of the beam spots on the mirrors 
and of the waist, as well as an angle deviation at the waist.
We introduce a pivot, where the non-congruent side of the aligned 
and the misaligned eigenmodes cross, indicated by the thick circle. 
We start with an arbitrary location of the pivot, and will shortly show
that it coincides with the bisecting point of the non-congruent side.

The angle of incidence on the flat mirrors changes by the same amount~%
$\theta$, but with opposite sign, resulting in the following changes of the 
large angles: $\gamma^{'}=\gamma+2\theta$ and $\gamma^{''}=\gamma-2\theta$.
Hence the small angle stays unchanged: $\phi^{'} = 2\phi$. 

\begin{figure}[htb]
\centerline{\includegraphics[width=100mm]{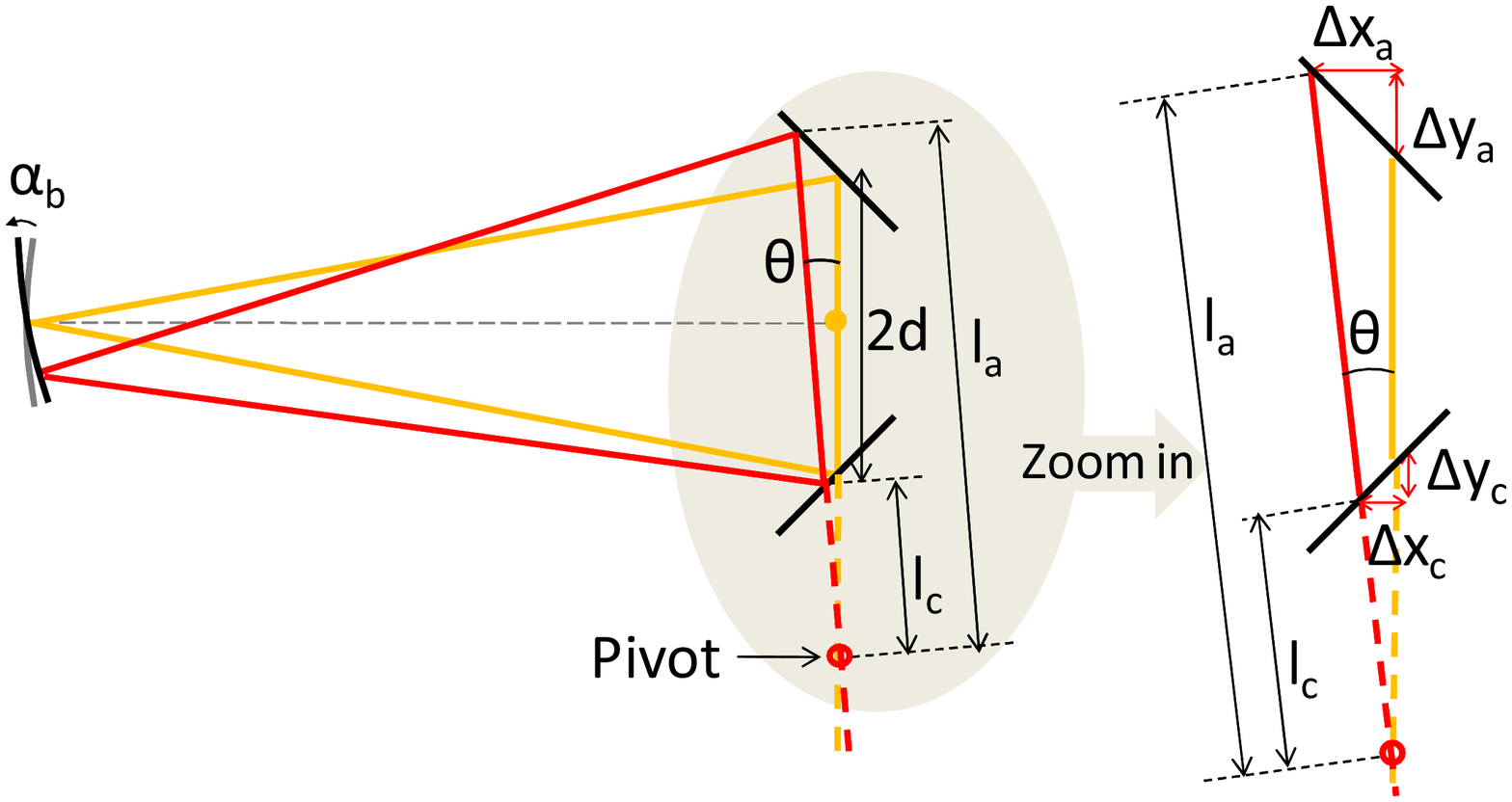}}
\caption{Closer view of the two flat mirrors and the pivot. It still shows the
hypothetical eigenmode where the pivot and the bisecting point do not match.
\label{005}}
\end{figure}

\begin{figure}[htb]
\centerline{\includegraphics[width=100mm]{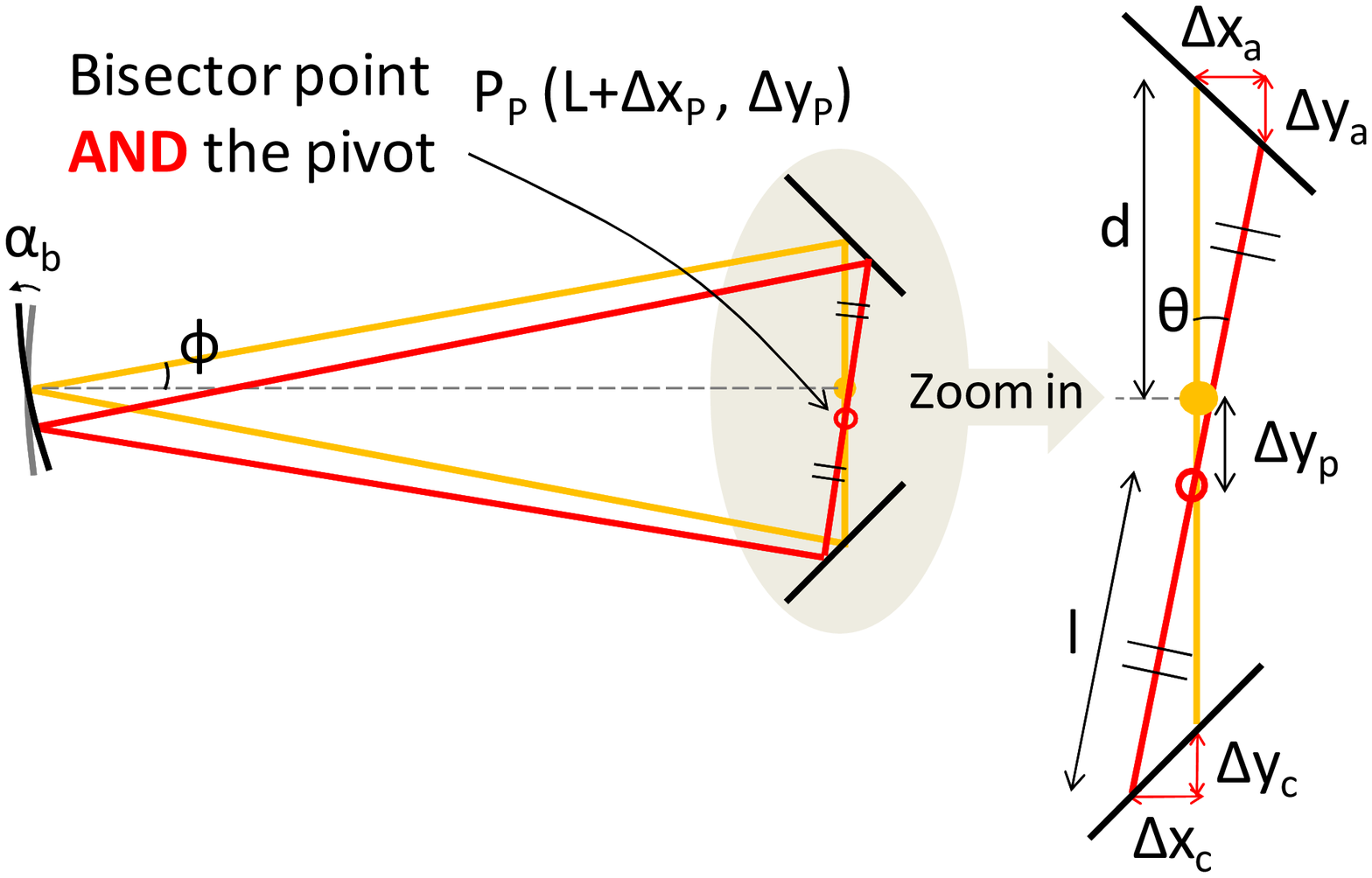}}
\caption{Closer view of the two flat mirrors. Here, the pivot and the bisecting 
point overlap, and the $y$ coordinate of the pivot is denoted by~$\Delta y_p$.
\label{006}}
\end{figure}

\begin{figure}[htb]
\centerline{\includegraphics[width=120mm]{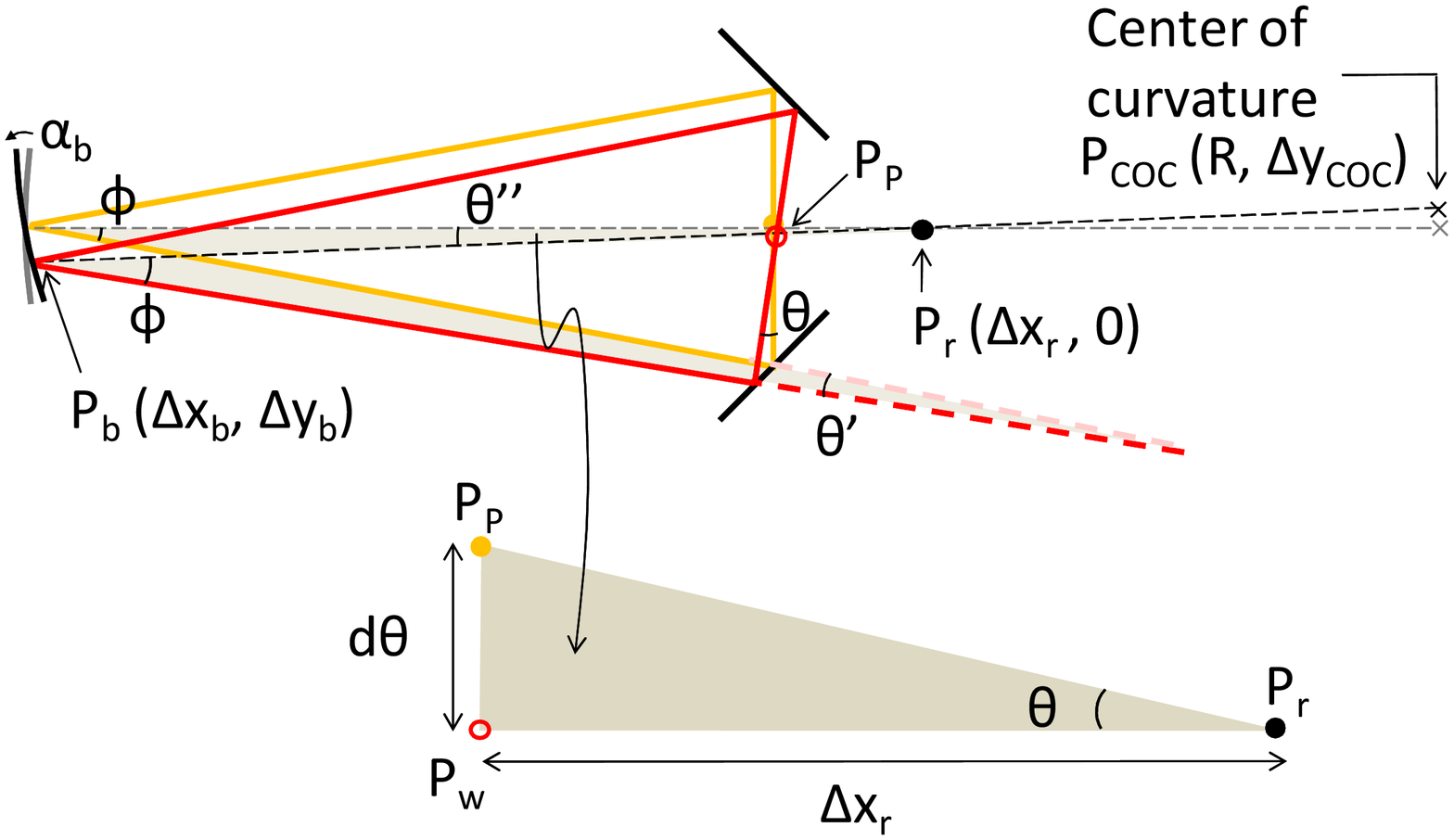}}
\caption{Radius vectors of the aligned and the misaligned cases. They cross at 
the point $P_r$, from which the angle deviation and the 
pivot location are calculated. \label{008}}
\end{figure}

\begin{figure}[htb]
\centerline{\includegraphics[width=120mm]{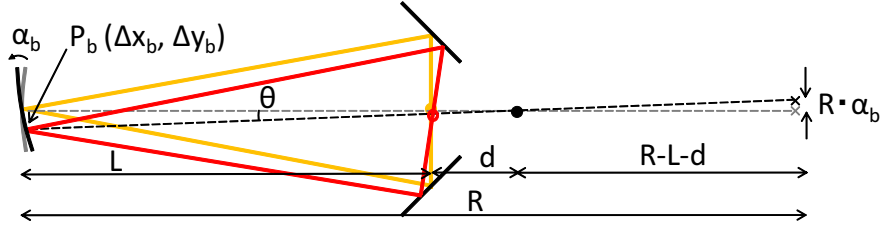}}
\caption{Length information needed to calculate $\theta$ and the spot position 
change on~$\rm M_b$. \label{009}}
\end{figure}

Looking at the flat mirrors, as shown in Fig.~\ref{005}, and applying the
approximation given by Equ.~\ref{gamap}, one sees that $\Delta x_a=\,-\Delta
y_a$ and  $\Delta x_c=\,\Delta y_c$. 
The following set of equations describe the shift of the spot positions:
\begin{equation}
\Delta x_a = l_a\sin{\theta}
\end{equation}
\vspace{-4mm}
\begin{equation}
\Delta x_c = l_c\sin{\theta}
\end{equation}
\vspace{-4mm}
\begin{eqnarray}\label{pivot}
2d \,&=\, l_a\cos{\theta}-l_c\cos{\theta}-\left|\Delta y_a\right|-\left|\Delta
y_c\right|\\
&=\,(l_a-l_c)\cos{\theta}-\left(\left|\Delta x_a\right|+\left|\Delta x_c\right|
\right)\\
&\approx\,l_a-l_c-\left(\left|l_a\right|+\left|l_c\right|\right)\,\theta
\nonumber
\end{eqnarray}
where $l_a$ and $l_c$ are the distances from the pivot to $P_a$ 
and $P_c$ along the beam, respectively. 
The left hand side of Equation~\ref{pivot} is constant, hence the right 
hand side must be independent of $\theta$, yielding the following relations:
\begin{equation}
l_a= - l_c
\end{equation} 
\begin{equation}
\left|l_a\right|=\, \left|l_c\right|=d/\cos{\theta}\equiv l
\end{equation} 
%
\begin{equation}
\Delta x_a=\, -\Delta x_c=\,-\Delta y_a=\, -\Delta y_c=\,l\,\sin{\theta}=\,d\,
\tan{\theta}\approx\,d\,\theta
\end{equation}
This automatically means that the pivot ($P_p$) bisects the non-congruent
side, as shown in Fig.~\ref{006}, where the changes in the location of the
pivot is denoted by $\Delta x_p$ and $\Delta y_p$.
It also shows the details around the flat mirrors, from which the pivot 
location with respect to the original waist is given by
\begin{equation}
\Delta x_p =O\left(\theta^2\right)=0
\end{equation} 
\begin{equation}
\Delta y_p =\,d-l\cos{\theta}+\Delta y_a=-d\,\theta
\end{equation}
Connecting the beam spot on the curved mirror ($P_b$) and the 
bisector point (the pivot, or $P_p$), one can see that it bisects 
the beam angle at $\rm M_b$ into $\phi$, as shown in Fig.~\ref{008}. 
This means that the line passes through the center of curvature, 
($P_{\rm COC}$), whose coordinate along the $y$-axis is given by 
\begin{equation}
\Delta y_{\rm COC}=\, R\cdot\alpha_b
\end{equation} 
Focusing on the shaded triangles shown in Fig.~\ref{008}, one can
see that $\theta^{'}=\,\theta$, and,
comparing the two triangles, one can also see that 
$\theta^{''}=\,\theta^{'}=\theta$. 
The radius vectors of the aligned and misaligned mirror, indicated by the dotted
lines in Fig.~\ref{008}, cross at point $P_r$. 
By focusing on the triangle consisting of the original waist 
($P_w$), the pivot, ($P_p$), and $P_r$, as shown in the lower 
triangle in Fig.~\ref{008}, one can see that the $x$ coordinate of the point 
$P_r$ is given by
\begin{equation}
\Delta x_r=\, d\,\theta/\tan{\theta}\approx\,d
\end{equation} 
Figure~\ref{009} lists all the length information that is needed to calculate 
the angle $\theta$ and the spot position on $\rm M_b$. 
These are given by the following set of equations:
\begin{equation}
\theta\,\approx\,\tan{\theta}=\, R\,\alpha_b/\left(R-L-d\right)
\end{equation}
\vspace{-4mm}
\begin{equation}
\Delta x_b =O\left(\Delta y_b^2\right)=0
\end{equation} 
\vspace{-4mm}
\begin{equation}
\Delta y_b=\,-\left(L+d\right)\,\tan{\theta}
\approx\,-\left(L+d\right)\theta 
=\,-R\,\alpha_b\cdot\left(L+d\right)/\left(R-L-d\right)
\end{equation}

\begin{figure}[htb]
\centerline{\includegraphics[width=120mm]{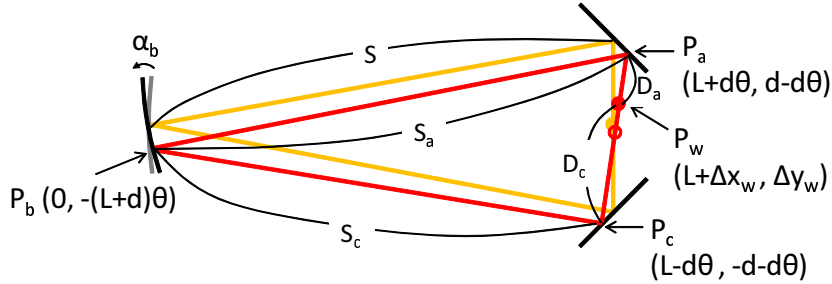}}
\caption{Locations of the new spot positions on the mirrors. By using them the 
new waist location is calculated. \label{010}}
\end{figure}

Having calculated the new spot positions on the mirrors, we now
calculate where the new waist is. 
In order for the wavefront curvature of the beam to match 
the radius of curvature of the curved mirror~$\rm M_b$, the path 
lengths from the waist to the mirror $\rm M_b$ via $\rm M_a$ and via 
$\rm M_c$ should be the same, i.e.\ in Fig.~\ref{010} it should be
\ $ S_a + D_a = S_c + D_c = S + d$\,. 
By calculating the distances $S_{a}$ and $S_b$ in the following equations,  
we also will obtain the distances $D_a$ and~$D_b$\,: 
\begin{eqnarray}\label{Sa}
S_{a}&=\,\left\{\left(L+d\,\theta\right)^2+\,\left(d+L\,\theta\right)^2\right\}
^{1/2}\\
&\approx\,\sqrt{L^2+d^2}\,\left(1+\frac{4Ld\,\theta}{L^{2}+d^{2}} \right)^{1/2}
  \qquad& \theta^2=0\nonumber\\
&\approx\,\sqrt{L^2+d^2}\,\left(1+\frac{2Ld\,\theta}{L^2+d^2}\right)\qquad&\frac
{2Ld\,\theta}{L^2+d^2} \ll 1\nonumber\\
&=\,S +2\,d\,\theta\qquad&d^2/L^2=0\nonumber
\end{eqnarray}
\begin{equation}\label{Da}
D_{a}=\,S+d-S_{a}=\,d-2\,d\,\theta
\end{equation} 
In a similar way
\begin{equation}\label{Sc}
S_{c}=\,\left\{\left(L-d\,\theta\right)^2+\,\left(-d+L\,\theta\right)^2\right\}
^{1/2}=\,S +2d\,\theta
\end{equation} 
\begin{equation}\label{Dc}
D_{c}=\,S+d-S_{b}=\,d+2d\,\theta
\end{equation}
Hence, the new waist location is given by the following:
\begin{equation}
\Delta x_w=\,O\left(\theta^2\right)=0
\end{equation}
\begin{equation}\label{dyw}
\Delta y_w=\,\left(d-d\,\theta-D_a\right)\cos{\theta}\approx\,d\,\theta =\, dR\,\alpha_b/\left(R-L-d\right)
\end{equation}

To summarize, a misalignment in $+(-)\alpha_b$ causes a clockwise 
(counter-clockwise) rotation of
the non-congruent side around the bisecting point, yielding the 
long sides to rotate synchronously. 
As a result all the beam spot positions change by the amounts
given by the radial distances with the bisecting point being 
the origin of the system of radial coordinates. 

\subsection{Misalignment in $\alpha_+$}
\begin{figure}[htb]
\centerline{\includegraphics[width=120mm]{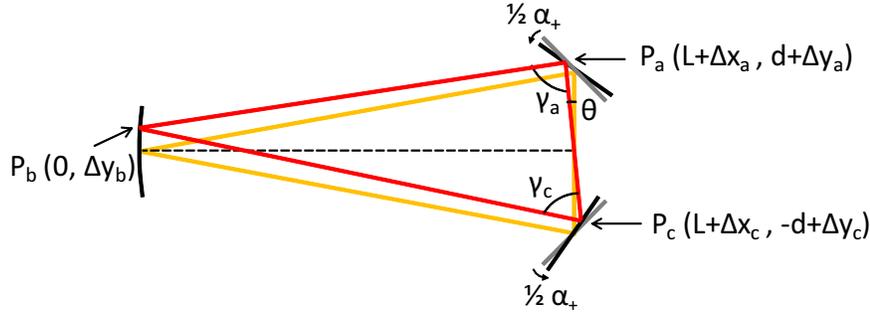}}
\caption{Cavity eigenmodes of the aligned and the misaligned ($\alpha_+$) cases.
 \label{011}}
\end{figure}

\begin{figure}[htb]
\centerline{\includegraphics[width=75mm]{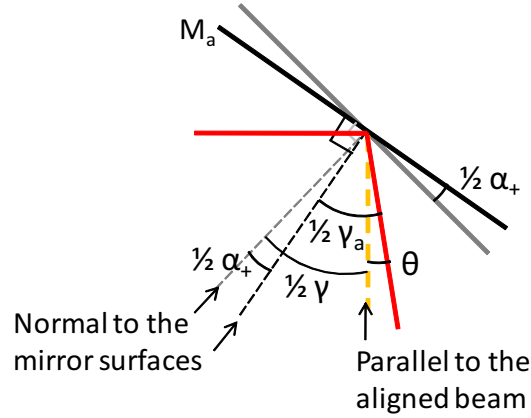}}
\caption{Closer view on the change in one of the larger angles, $\gamma_a$.
\label{012}}
\end{figure}

\begin{figure}[htb]
\centerline{\includegraphics[width=120mm]{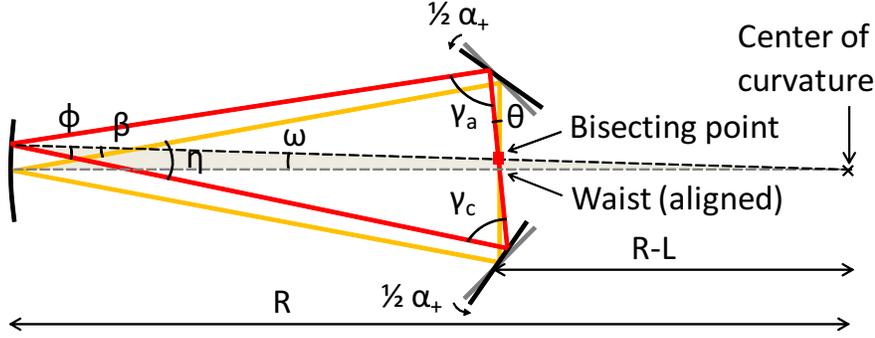}}
\caption{Ancillary angles: $\beta$, $\eta$ and $\omega$, which are use to 
calculate $\theta$.\label{014}}
\end{figure}

\begin{figure}[htb]
\centerline{\includegraphics[width=70mm]{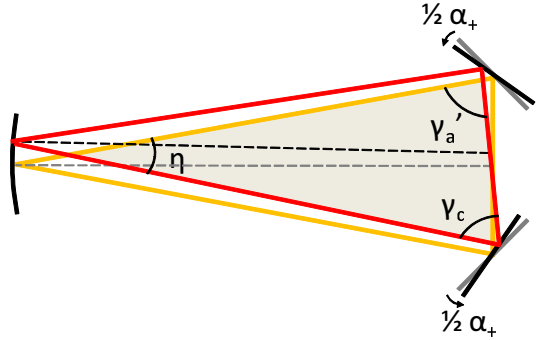}}
\caption{Yet another ancillary angle $\gamma_a^{'}$ to calculate $\eta$.
\label{015}}
\end{figure}

In the case of $\alpha_+\neq 0$, there is no obvious symmetry line, 
thus we will start from a general case. 
Changes on the spot positions, the beam angle at the waist, and the two larger 
angles are 
defined as shown in Fig.~\ref{011}. 
Figure~\ref {012} focuses on the beam angle change on mirror~$\rm M_{a}$. 
Drawing helping lines such as the one that is parallel to the aligned 
beam (indicated by the light colored thick dotted line), 
as well as lines that are normal to both the aligned and the 
misaligned mirror surfaces (indicated by the light thin, 
and dark thin dotted lines, respectively) one can see that 
half of $\gamma_a$ is given by 
$\gamma_{a}/2\,=\gamma/2+\theta-\alpha_{+}/2$.
Hence,
\begin{eqnarray}
&&\gamma_a=\,\gamma + \left(2\theta-\alpha_+\right)
\end{eqnarray}
In a similar manner, $\gamma_c$ is given by
\begin{eqnarray}
&&\gamma_c=\,\gamma - \left(2\theta-\alpha_+\right)
\end{eqnarray}
This means that the sum of the two angles stays unchanged, 
yielding no change in the small angle~$\phi$. 
Then the line that connects $P_b$ with the center of curvature of $\rm M_b$
(from here on this is called the \textsl{radius}), should bisect the short side,
due to the fact that~$d\ll L$. 
The bisecting point is indicated by the square point in Fig.~\ref{014}.
Here, we introduce some ancillary angles $\beta$ and $\eta$, together with 
$\omega$,
which is the angle of the radius with respect to the aligned case.  
Focusing on the shaded area, one can see that the ancillary angles are given by
\begin{eqnarray}
\beta&=\,\phi+\omega \qquad\rm and\\
\eta&=\,\phi+\beta=\,2\phi+\omega\label{eta1}
\end{eqnarray}
$\eta$ can be expressed using $\gamma$ if one focuses on the shaded triangle
shown in Fig.~\ref{015}, 
introducing a new ancillary angle $\gamma_a^{'}=\,\gamma+\theta$,
and it is given by
\begin{equation}\label{eta2}
\eta=\,\pi-\left(\gamma_{a}^{'} + \gamma_c\right)
=\pi-\left\{\left(\gamma+\theta\right)+\gamma-\left(2\theta - \alpha_+\right)
\right\}
=\pi-2\gamma+\theta-\alpha_+
\end{equation}
By comparing Equations~\ref{eta1} and~\ref{eta2} the angle $\omega$ 
is given by the following equations:
\begin{equation}
\pi-2\gamma+\theta-\alpha_{+}=\,2\phi+\omega
\end{equation}
\begin{equation}\label{omega1}
\omega=\,\pi-\left(2\gamma+2\phi\right)+\theta-\alpha_{+}\nonumber=\,\theta-
\alpha_{+}
\end{equation}

\begin{figure}[htb]
\centerline{\includegraphics[height=80mm]{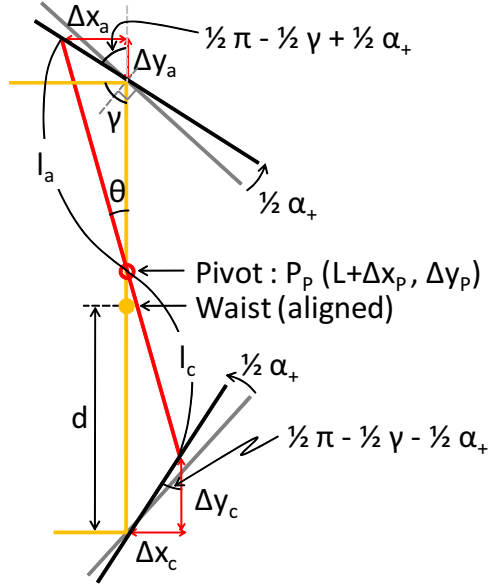}}
\caption{Length relations around the flat mirrors. From this the lengths
$l_{a}$ and $l_{c}$ from the pivot to the beam spots on the two mirrors
are calculated.)\label{016}}
\end{figure}
In order to gain additional information to finally calculate $\theta$, 
we focus on some lengths as shown in Fig.~\ref{016}. 
The pivot ($P_p$) is indicated by the thick circle and changes in its location
 are denoted by $\Delta x_p$ and~$\Delta y_p$, and the two lengths 
from the pivot to the two beam spots by $l_a$ and~$l_b$.
Changes in the coordinates of the beam spot position on $\rm M_a$ are given by 
the following equations:
\begin{equation}\label{dxa}
\Delta x_{a}=\,-l_{a}\sin{\theta}\,\approx\,-l_{a}\,\theta 
\end{equation}
\begin{equation}\label{dya}
\Delta y_{a}=\,\frac{-\Delta x_{a}}
{\tan{\left(\pi/2-\gamma/2+\alpha_+/2\right)}}\approx\,\frac{1-\alpha_{+}/2}
{1+\alpha_{+}/2}\,l_{a}\,\theta\approx\,\left(1-\alpha_{+}\right)l_{a}\,\theta
\end{equation}
and $\Delta x_c$ and $\Delta y_c$ by:
\begin{equation}\label{dxc}
\Delta x_{c}=\,l_{c}\sin{\theta}\,\approx\,l_{c}\,\theta
\end{equation}
\begin{equation}\label{dyc}
\Delta y_{c}=\,\frac{\Delta x_{c}}
{\tan{\left(\pi/2-\gamma/2-\alpha_+/2\right)}}\approx\,\frac{1+\alpha_{+}/2}{1-
\alpha_{+}/2}\,l_{c}\,\theta\approx\,\left(1+\alpha_{+}\right)l_{c}\,\theta
\end{equation}
The length of the non-congruent side is then expressed by the following:
%
\begin{equation}\label{2d2}
2d=\,l_{a}\,\cos{\theta}+\Delta y_{a}+l_{c}\,\cos{\theta}+\Delta y_{c}
\approx\,l_{a}+l_{c}-\left\{l_{a}-l_{c}-\left(l_{a}+l_{c}\right)
 \alpha_{+}\right\}\theta 
\end{equation}
Since the left-hand side of Equation~\ref{2d2} does not depend 
on the misalignment angle $\theta$, the angle dependent term of the 
right-hand side should be zero, hence, 
\begin{equation}\label{lalc}
  l_{a}-l_{c}-\left(l_{a}+l_{c}\right)\alpha_{+}=\,0
\end{equation}
From Equations~\ref{2d2} and~\ref{lalc} the following relations 
can be obtained:
\begin{eqnarray}
l_{a}&=d\left(1+\alpha_+\right)\\
l_{c}&=d\left(1-\alpha_+\right)
\end{eqnarray}
With this knowledge we can calculate the location of the pivot in the 
following way:
\begin{equation}
\Delta x_p =\,O\left(\theta^2\right)=0
\end{equation}
\vspace{-4mm}
\begin{equation}
\Delta y_{p}=l_{c}\,\cos{\theta}+\Delta y_c-d\approx\,d\left(\theta-\alpha_{+}
\right)
\end{equation}
\begin{figure}[htb]
\centerline{\includegraphics[height=80mm]{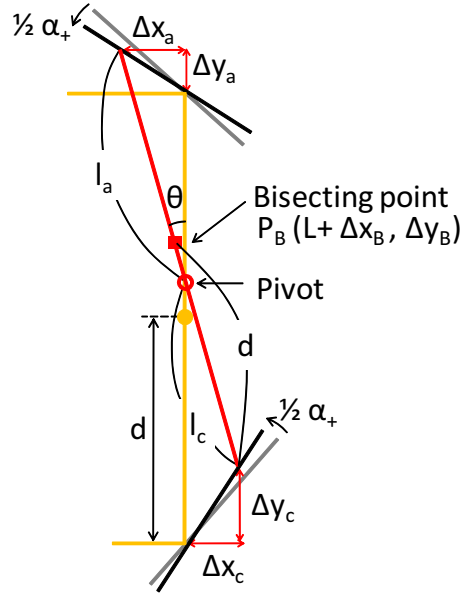}}
\caption{Length relations around the flat mirrors, including the pivot location.
 From this the spot position changes on the flat mirrors are calculated.
\label{017}}
\end{figure}
The location of the bisecting point, as shown in Fig.~\ref{017}, 
can be calculated in a similar way, and the coordinates are given by 
\begin{equation}
\Delta x_B=O\left(\theta^2\right)=0
\end{equation}
\vspace{-4mm}
\begin{equation}
\Delta y_{B}=d\,\cos{\theta}+\Delta y_{c}-d
\approx\,\left(1+\alpha_+\right)\left(1-\alpha_+\right)d\,\theta
\approx\,d\,\theta
\end{equation}
Then, focusing on the triangle that consists of the center of curvature, 
the waist (in the aligned case), and the bisecting point (indicated by the 
right part of the shaded area in Fig.~\ref{014}), one can obtain another 
relation for $\omega$ and $\theta$ which is given by
\begin{equation}\label{omega2}
\omega\approx\,\tan{\omega}=d\,\theta/\left(R-L\right)
\end{equation}
From Equations~\ref{omega1} and~\ref{omega2} one can finally obtain the 
relation between $\theta$ and $\alpha_{+}$\,:
\begin{eqnarray}
\theta = \frac{R-L}{R-L-d}\cdot \alpha_{+}
\end{eqnarray}
Using $\theta$, the spot positions on the three mirrors (see Equations 
\ref{dxa}, \ref{dya}, \ref{dxc}, and \ref{dyc}) can further be calculated. 
This yields the following equations:
\begin{equation}
\Delta x_{a}=\,-d\left(1+\alpha_+\right)\theta\approx\,-\frac{d\left(R-L\right)}
{R-L-d}\cdot \alpha_{+}
\end{equation}
\vspace{-2mm}
\begin{equation}
\Delta y_a =\,\left(1-\alpha_+\right)\Delta x_a\approx\,-\frac{d\left(R-L\right)
}{R-L-d}\cdot \alpha_{+}
\end{equation}
And in similar ways, 
\begin{eqnarray}
\Delta x_{c}&=\,\frac{d\left(R-L\right)}{R-L-d}\cdot \alpha_{+}\\
\Delta y_{c}&=\,\frac{d\left(R-L\right)}{R-L-d}\cdot \alpha_{+}
\end{eqnarray}
and 
\begin{equation}
\Delta x_b=\,O\left(\Delta y_b^2\right)=\,0
\end{equation}
\vspace{-5mm}
\begin{equation}
\Delta y_{b}=\,R\cdot\omega=\,R\cdot\left(\theta-\alpha_{+}\right)
=\,\frac{d\,R}{R-L-d}\cdot\alpha_{+}
\end{equation}
Then the waist location can be calculated in the same way 
as shown in equations \ref{Sa} to~\ref{dyw}, 
and the following can be shown:
\begin{equation}\label{Sa2}
S_{a}=\,\left\{\left(L+\Delta x_a\right)^2+\,\left(d+\Delta y_a -\Delta y_b\,
\right)^2\right\}^{1/2}\approx\,S-\left(d\,\theta+\Delta y_p\right)
\end{equation}
\vspace{-3mm}
\begin{equation}\label{Da2}
D_{a}=\,d+d\,\theta+\Delta y_p
\end{equation}
In a similar way we obtain
\begin{equation}\label{Sb2}
S_{b}=\,S+\left(d\,\theta+\Delta y_p\right)
\end{equation}
\vspace{-4mm}
\begin{equation}\label{Db2}
D_{b}=\,d-d\,\theta+\Delta y_p
\end{equation}
Therefore the new waist location is given by
\begin{equation}\label{wxb}
\Delta x_{w}=\,O\left(\theta^2\right)=0
\end{equation}
\vspace{-6mm}
\begin{equation}\label{wyb}
\Delta y_{w}=\,\left(d+\Delta y_a-D_a\right)\cos{\theta}\,\approx\,-\Delta y_p =
-\frac{d^2}{R-L-d}\cdot \alpha_{+}
\end{equation}

To summarize, a misalignment in $+(-)\alpha_+$ causes a counter-clockwise
 (clockwise) rotation of the non-congruent side around a point 
 that does not coincide with the bisecting point.
This yields a clockwise (counter-clockwise) rotation
$\omega$ (which is very small compared to the 
misalignment angle~$\alpha_+$ ) of the geometrical axis of a corner 
reflector consisting of the two flat mirrors. 
As a result, the eigenmode changes in a ``non uniform'' way, with
each spot position change being 
smaller than the misalignment case of $\alpha_b$.

\section{Vertical misalignments}
\begin{figure}[htb]
\centerline{\includegraphics[width=120mm]{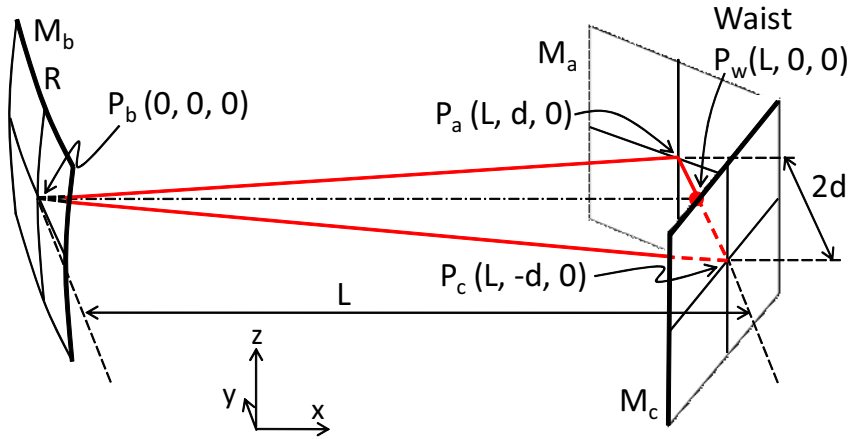}}
\caption{\label{018}3D view of a triangular cavity. $\rm M_a$ and $\rm M_c$ are 
the flat mirrors, and $\rm M_b$ has a radius of curvature of~$R$. The positions 
where the beam hits the mirror $\rm M_i$ are denoted by~$P_i$.}
\end{figure}

When considering vertical misalignments, it is necessary to view 
the cavity as a 3D body, as shown in Fig.~\ref{018}. 
Notations of all the properties are the same as that shown 
in Fig.~\ref{000}.
\subsection{Misalignment in $\beta_b$}
\begin{figure}[htb]
\centerline{\includegraphics[width=120mm]{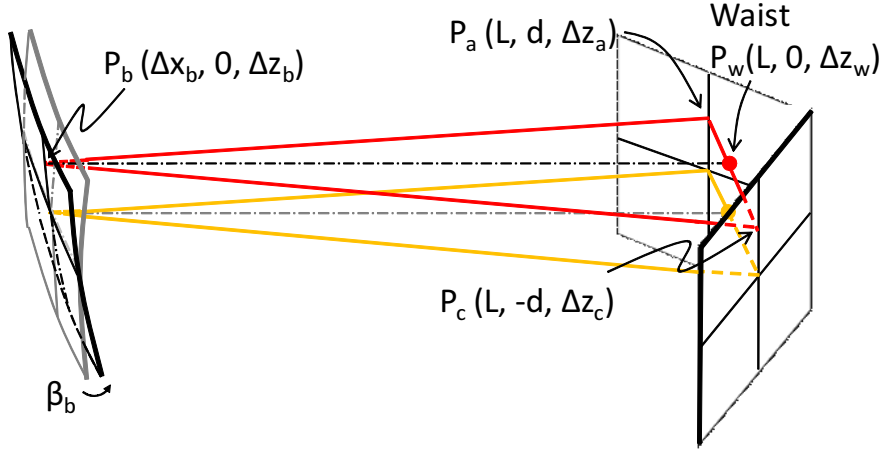}}
\caption{Cavity eigenmodes of the aligned (lighter colored triangle) and the 
misaligned by $\beta_b$ (darker colored triangle) cases. This type of 
misalignment does not affect the mirror alignment in $y$-direction, hence the 
eigenmode only changes along the $z$-axis.\label{019}}
\end{figure}

\begin{figure}[htb]
\centerline{\includegraphics[width=120mm]{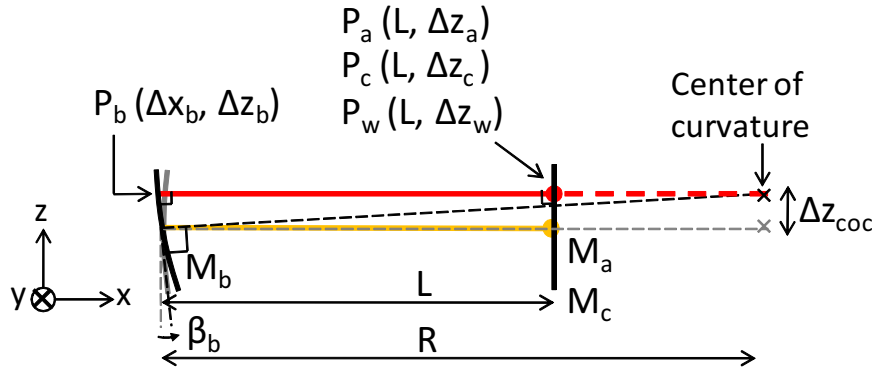}}
\caption{Projection of the triangular cavity onto the $x-z$ plane. It allows one
 to view the cavity as a plane cavity. The eigenmode is defined by the line 
that is orthogonal to the flat mirror and passes through the center of 
curvature.\label{020}}
\end{figure}

A misalignment around the $y$-axis by $\beta_b$, as shown in 
Fig.~\ref{019}, does not affect the mirror alignment in $y$-direction, 
hence there is no change in eigenmode in that direction. 
Then it is possible to project the cavity onto the $x-z$ plane for simplicity, 
as shown in Fig.~\ref{020}, and treat it as a plane cavity.
The eigenmode of the cavity is defined by the line that is orthogonal to the 
flat mirrors and passes through the center of curvature, as described in
~\cite {Gerhard}.
It is obvious that the eigenmode is also orthogonal to the curved mirror, 
yielding the shifts in $z$-direction of all of the spot positions to have the 
same size. 
The normal vector on the mirror $\rm M_b$ is tilted by~$\beta_b$, 
hence the center of curvature, whose $z$ coordinate is denoted by $z_{\rm coc}$,
shifts by \,$\Delta z_{\rm coc}=\beta_b \cdot R$\,. 
Therefore we have the following relations:
\begin{eqnarray}
\Delta x_{a}=O\left(\beta_{b}^2\right)=0\\
\Delta z_{a}=\,\Delta z_{b}=\,\Delta z_{c}=\,\Delta z_{w}=\,\Delta 
  z_{\rm coc}=\,\beta_{b} \cdot R
\end{eqnarray}

To summarize, a misalignment in $+(-)\beta_b$ causes an upward (downward)
shift of the center of curvature along the $z$-axis, yielding a synchronous
shift of the plane of the eigenmode by an amount 
proportional to the radius of curvature of the curved mirror.

\subsection{Misalignment in $\beta_{+}$}
\begin{figure}[htb]
\centerline{\includegraphics[width=120mm]{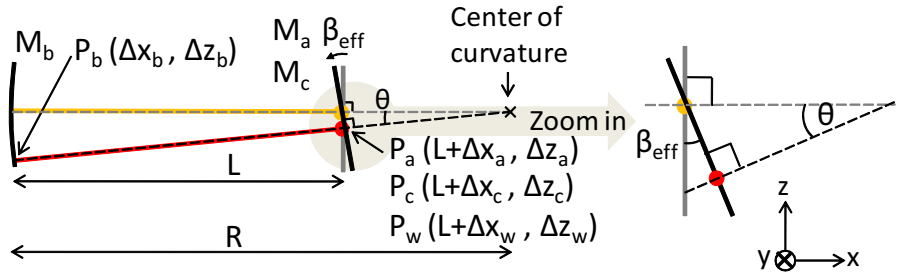}}
\caption{Projection of the triangular cavity onto the $x-z$ plane. It allows one
 to view the cavity as a plane cavity. The eigenmode is defined by the line 
that is orthogonal to the flat mirrors and passes through the center of 
curvature. In the right part, an enlarged cut-out around one flat mirror is 
shown.\label{022}}
\end{figure}

Similar to $\beta_{b}$, $\beta_{+}$ has no effects in $y$-direction. 
However, since the $y_a$-axis and the $y_c$-axis are rotated by 
$\pm \left(\frac{\pi}{2}-\frac{\gamma}{2}\right)\approx\,\pm \frac{\pi}{4}$
around the $y$-axis, respectively, the projection of a misalignment by 
$\beta_{+}/2$ around the two axes becomes $\frac{1}{2}\,\beta_+/\sqrt{2}$.
Section~4.10, (page 100-102) of reference~\cite{Freise}, gives a detailed 
explanation of this effect by using a vector algebra and we will not describe it
 in this paper. 
\begin{figure}[htb]
\centerline{\includegraphics[width=120mm]{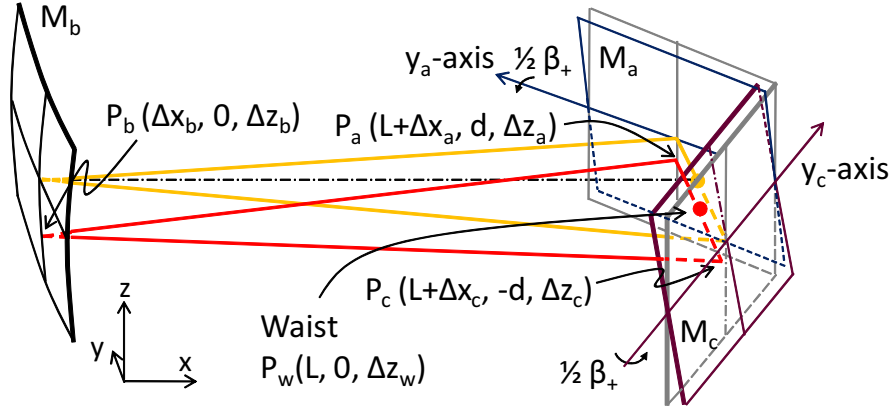}}
\caption{Cavity eigenmodes of the aligned and the misaligned ($\beta_{+}$) 
cases. This type of misalignment does not affect the mirror alignment in 
$y$-direction, hence the eigenmode only changes along the $z$-axis.\label{021}}
\end{figure}
For convenience we introduce an effective misalignment angle
$\beta_{\rm eff}=\beta_{+}/\sqrt{2}$. 
The projection of the flat mirrors are rotated by $\beta_{\rm eff}/2$ around the
 $y$-axis and the effect is doubled because of the two reflections, hence, seen 
as a plane cavity, the misalignment angle is given by~$\beta_{\rm eff}$, as 
shown in Fig.~\ref{022}.
The eigenmode of this cavity is defined by  the line that passes through 
the center of curvature and intersects 
the flat mirrors orthogonally, as described in~\cite{Gerhard}. 
The angle formed by the eigenmodes of the aligned and misaligned cases is 
denoted by $\theta$ in Fig.~\ref{022}, and it becomes obvious that $\theta=\,
\beta_{\rm eff}$ when one focuses around the area of the flat mirrors, 
as shown in the enlarged cut-out in the right part of Fig~\ref{022}.
Therefore the following equations yield the spot position changes:
\begin{equation}
\Delta x_{a,\, b,\, c,\, {\rm and}\, w}=\,O\left(\beta_{+}^2\right)=\,0
\end{equation}
\vspace{-4mm}
\begin{equation}
\Delta z_{a}=\,\Delta z_{c}=\,\Delta z_{w}=\,\beta_{\rm eff}
  \cdot\left(R-L\right)=\beta_{+}\cdot\left(R-L\right)/\sqrt{2}
\end{equation}
\vspace{-4mm}
\begin{equation}
\Delta z_{b}=\,\beta_{\rm eff}\cdot R=\,\beta_{+}\cdot R/\sqrt{2}
\end{equation}

To summarize, a misalignment in $+(-)\beta_+$ causes a
counter-clockwise (clockwise) tilt of the geometrical axis 
of the two flat mirrors around the center of curvature.
As a result the plane of the eigenmode tilts synchronously.

\subsection{Misalignment in $\beta_{-}$}
\begin{figure}[htb]
\centerline{\includegraphics[width=120mm]{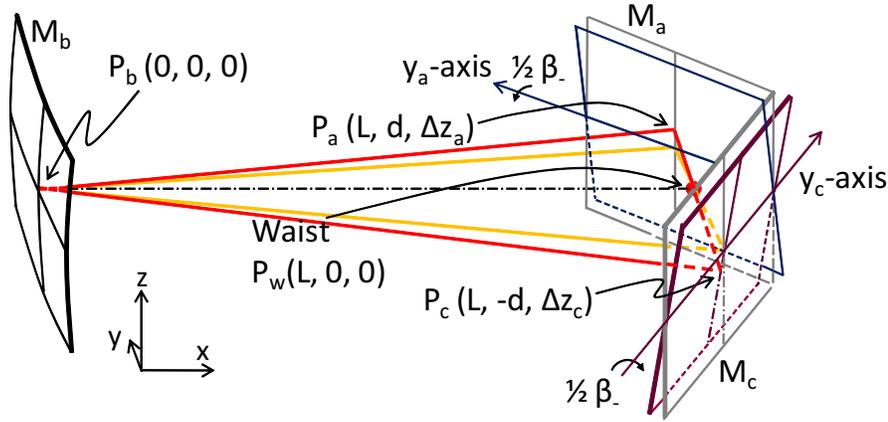}}
\caption{Cavity eigenmodes of the aligned and the misaligned ($\beta_{-}$) 
cases. The beam spot position and the waist position stays unchanged.
\label{023}}
\end{figure}

\begin{figure}[htb]
\centerline{\includegraphics[width=100mm]{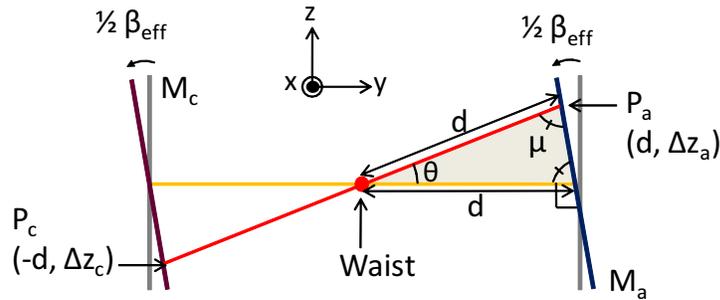}}
\caption{Projection of the triangular cavity onto the $y-z$ plane. The plane of 
the cavity is rotated around the $x$-axis by~$\theta$, however the lengths of all
 the sides of the triangle remain unchanged.\label{024}}
\end{figure}

Here, mirrors $\rm M_a$ and $\rm M_c$ rotate around the $y_a$ and $y_c$-axis by 
$\pm 1/2\,\beta_{-}$, respectively, as shown in Fig.~\ref{023}. 
When the two opposite misalignment angles on mirrors $\rm M_a$ and $\rm M_c$ are
 projected onto the $x-z$ plane, they appear as rotations around the $y$-axis 
by~$\pm \frac{\beta_{\rm eff}}{2}$, respectively, yielding no change along the 
$z$-axis on the curved mirror~$\rm M_b$. 
On the other hand, when they are projected onto the $y-z$ plane, as shown in 
Fig.~\ref{024}, they both appear as rotations around the $z$-axis 
by~$\beta_{\rm eff}/2$, yielding  
shifts along the $z$-axis in the beam spot positions on the two flat 
mirrors by the same amount, but with opposite sign. 
Note that here $\beta_{\rm eff}\equiv\beta_{-}/\sqrt{2}$.
These spot position changes are symmetrical along the $y$-axis, thus they do not
 yield a change in the beam spot position on the curved mirror along the 
$y$-axis, nor
a change in the waist position (which is equidistant from the two spot 
positions) along the $y$-axis and $x$-axis. Hence the spot on the curved mirror and 
the waist remain unchanged, indicating that the new eigenmode is formed by 
rotating the aligned eigenmode around the $x$-axis by~$\theta$, yielding no
 change in the lengths on any sides of the triangle.

The inclination angle of the beam between the two flat mirrors with respect to 
the $x-y$ plane is denoted by $\theta$ in Fig.~\ref{024}. Focusing on the 
isosceles triangle as indicated by the shaded triangle in the figure, whose 
equal angles are denoted as $\mu$, the inclination angle is given by the 
following equations:
\begin{eqnarray}
\mu=\,\pi/2-\beta_{\rm eff}/2\\
\theta=\,\pi-2\mu=\,\beta_{\rm eff}
\end{eqnarray}
Therefore the beam spot position shifts on the two mirrors are calculated to be
\begin{equation}
\Delta z_{a}=\,-\,\Delta z_{c}=\,d\cdot \alpha_{\rm eff} =\,d\cdot
  \beta_{-}/\sqrt{2}
\end{equation}

To summarize, a misalignment in $+(-)\beta_-$ causes no
change in the spot position on the curved mirror and a counter-clockwise
(clockwise) rotation of the non-congruent side around the $x$-axis. 
As a result, the plane of the eigenmode rotates synchronously.

\section{Result and comparison}
Tables \ref{horiz} and \ref{vert} show the results from the geometrical analysis,
 and compare them to the simulation results obtained by using two simulation 
tools. 
One is \textsc{OptoCad}~\cite{oc} and the other is \textsc{Ifocad}~\cite{ic}.
We used them to trace the Gaussian beam through our 
triangular cavity model that has the design parameters for the AEI~10\,m 
Prototype reference cavity. These parameters are given as follows: $R=37.8\,m$, 
$L=10.05\,m$, and $d=0.15\,m$ . By inserting these values into our 
geometrical model, we obtained the corresponding numerical values.
Due to the fact that O{\scriptsize PTO}C{\scriptsize AD} 
is 2-dimensional we only used it for simulating the horizontal misalignment 
types. 
\begin{table}
\caption{\label{horiz}Horizontal misalignment comparison.}
\small
\begin{tabular*}{\textwidth}{@{}l*{10}{@{\extracolsep{6pt plus
12pt}}r}}
\br
Type&Method&\centre{2}{$\rm\ \quad M_a$}&\centre{2}{$\rm\ \quad M_c$}&\centre{2}
{$\rm\ M_b$}&\centre{3}{$\rm Waist$}\\
\ns
&&$\Delta x$&$\Delta y$&$\Delta x$&$\Delta y$&$\Delta x$&$\Delta y$&$\Delta x$
&$\Delta y$& $\rm Angle$\\
\mr
&Geom.Analy.&0.205&-0.205&-0.205&-0.205&0&-13.970&0&0.205&-1.370\\
$\alpha_b$&\textsc{OptoCad}&0.206&-0.202&-0.206&-0.202&0&-13.974&0&0.209&-1.370\\
&\textsc{Ifocad}&0.205&-0.202&-0.205&-0.202&0&-13.974&0&0.209&-1.370\\
\mr
&Geom.Analy.&-10.051&10.051&-10.051&-10.051&0&0&-10.051&0&0\\
$\alpha_-$&\textsc{OptoCad}&-10.051&9.902&-10.051&-9.902&0&0&-10.051&0&0\\
&\textsc{Ifocad}&-10.051&9.903&-10.051&-9.903&0&0&-9.9022&0&0\\
\mr
&Geom.Analy.&-0.151&0.151&0.151&0.151&0&0.205&0&-0.001&1.005\\
$\alpha_+$&\textsc{OptoCad}&-0.151&0.149&0.151&0.149&0&0.206&0&-0.003&1.005\\
&\textsc{Ifocad}&-0.151&0.149&0.151&0.149&0&0.205&0&-0.003&1.005\\
\br
\end{tabular*}
%
\caption{\label{vert}Vertical misalignment comparison.}
\small
\begin{tabular*}{\textwidth}{@{}l*{10}{@{\extracolsep{6pt plus
12pt}}r}}
\br
Type&Method&\centre{2}{$\rm\ M_a$}&\centre{2}{$\rm M_c$}&\centre{2}{$\rm\ M_b$}
&\centre{3}{$\rm Waist\ $}\\
\ns
&&$\Delta x$&$\Delta z$&$\Delta x$&$\Delta z$&$\Delta x$&$\Delta z$&$\Delta x$
&$\Delta z$& $\rm Angle$\\
\mr
$\beta_b$&Geom.Analy.&0&37.800&0&37.800&0&37.800&0&37.800&0\\
&\textsc{Ifocad}&0&37.800&0&37.800&0&37.800&0&37.800&0\\
\mr
$\beta_-$&Geom.Analy.&0&0.106&0&-0.106&0&0&0&0&0.707\\
&\textsc{Ifocad}&0&0.106&0&-0.106&0&0&0&0&0.702\\
\mr
$\beta_+$&Geom.Analy.&0&-19.622&0&-19.622&0&-26.729&0&-19.622&0\\
&\textsc{Ifocad}&0&-19.770&0&-19.770&0&-26.930&0&-19.770&0\\
\br
\end{tabular*}
\end{table}
\section{Conclusion}
The results of the geometrical analysis are in excellent agreement with the simulation 
results, showing sufficient accuracy for the design of an alignment control 
system for a triangular cavity. 
We have checked that all the discrepancies between the geometrical analysis and 
the simulations decrease by assigning real values for the two 
larger angles $\gamma$ to the geometrical analysis, instead of using~$\gamma=\,\pi/2$. 
This analysis can easily be extended to a cavity with more 
general shape if one follows the equations derived in this paper and modifies the 
method of approximation properly. 
The geometrical analysis not only serves as a method of checking a simulation 
result, but also gives an intuitive and handy tool to visualize the eigenmode 
of a misaligned triangular cavity.
%

\section{Acknowledgments}
This work was supported by the QUEST cluster of excellence of the Leibniz
Universit\"at Hannover. We like to thank Gerhard Heinzel for very helpful 
discussions on simulation challenges.


\section*{References}
\begin{thebibliography}{21}
\frenchspacing
%
%
%
%
\bibitem{Gerhard}
Heinzel~G 1999  Advanced optical techniques for laser-interferometric
gravitational-wave detectors, \emph{PhD thesis}\quad
{\tt ftp.rzg.mpg.de/pub/grav/ghh/ghhthesis.pdf}

\bibitem{Fumiko} Kawazoe~F \emph{et al.} 2010 
Designs of the frequency reference cavity for the AEI~10\,m prototype 
interferometer, 
\emph{J.\,Phys.\,Conf.\,Ser.} textbf{228} 012028

\bibitem{Freise}
Freise~A 2010  Frequency domain INterfErometer Simulation SotfwarE, 
\emph{FINESSE manual}\\
{\tt http://www.gwoptics.org/finesse/download/Finesse-0.99.8.pdf}

\bibitem{oc}
\textsc{OptoCad}\,(0.90c) 2010 
A Fortran~95 module for tracing Gaussian \textsf{TEM}$_{00}$ beams 
through an optical set-up, written by Roland Schilling\\
\emph{Simulation tool}\quad
{\tt http://www.rzg.mpg.de/\string~ros/optocad.html}

\bibitem{ic}
\textsc{Ifocad}\,2010 
A framework of C subroutines to plan and optimize the geometry
of laser interferometers, written by Gerhard Heinzel\\
\emph{Simulation tool}\\
\end{thebibliography}
\end{document}